\documentclass[twocolumn]{aastex701}

\newcommand\Lumpford{HD 101452}

\usepackage{upgreek}
\usepackage[T1]{fontenc}
\usepackage{subcaption}
\usepackage{graphicx}

\shorttitle{MIRI imaging of HD 101452}
\shortauthors{Bowens-Rubin et al.}

\begin{document}

\title{
A Binary Flux Calibrator Reveals the Scientific Potential of Short-integration JWST MIRI Imaging to Directly Detect sub-Jupiter Exoplanets}
\correspondingauthor{Rachel Bowens-Rubin}
\email{rbowens-rubin@mit.edu}

\author[0000-0001-5831-9530]
{Rachel Bowens-Rubin}
\email{rbowens-rubin@mit.edu}
\affiliation{Eureka Scientific Inc., 2542 Delmar Ave., Suite 100, Oakland, CA 94602, USA}

\author[0000-0002-9521-9798]
{Mary Anne Limbach}
\affiliation{Department of Astronomy, University of Michigan, Ann Arbor, MI 48109, USA}
\email{mlimbach@umich.edu}

\author[0000-0002-8400-1646]{Alexander Venner}
\affiliation{Max Planck Institute for Astronomy, 69117 Heidelberg, Germany}
\email{alexandervenner@gmail.com}

\author[0000-0003-4557-414X]
{Kyle Franson}
\affiliation{Astronomy Department, University of California Santa Cruz, 1156 High St, Santa Cruz, CA 95064, USA}
\altaffiliation{NHFP Sagan Fellow}
\email{kfranson@ucsc.edu}

\author[0000-0002-1533-9029]
{Emily K. Pass}
\affiliation{Kavli Institute for Astrophysics and Space Research, Massachusetts Institute of Technology, Cambridge, MA 02139, USA}
\email{}

\author[0000-0001-7246-5438]
{Andrew Vanderburg}
\affiliation{Center for Astrophysics $\vert$ Harvard \& Smithsonian, 60 Garden Street, Cambridge, MA 02138, USA}
\email{}

\author[0000-0002-7352-7941]{Kevin B. Stevenson}
\email{kevin.stevenson@jhuapl.edu}
\affiliation{Johns Hopkins APL, 11100 Johns Hopkins Rd, Laurel, MD 20723, USA}

\author[0000-0003-3904-7378]
{Logan Pearce}
\affiliation{Department of Astronomy, University of Michigan, Ann Arbor, MI 48109, USA}
\email{lapearce@umich.edu}

\author[0000-0001-8103-5499]
{Taylor L. Tobin}
\affiliation{Department of Astronomy, University of Michigan, Ann Arbor, MI 48109, USA}
\email{tltobin@umich.edu}

\author[0000-0003-0593-1560]
{Elisabeth Matthews}
\affiliation{Max-Planck-Institut für Astronomie, Königstuhl 17, D-69117 Heidelberg, Germany}
\email{matthews@mpia.de}

\author[0009-0004-0868-1186]
{Aiza Kenzhebekova}
\affiliation{Institute for Astronomy, University of Edinburgh, Royal Observatory, Blackford Hill, Edinburgh, EH9 3HJ, UK}
\email{A.Kenzhebekova@sms.ed.ac.uk}

\author[0000-0003-4614-7035]
{Beth Biller}
\affiliation{Institute for Astronomy, University of Edinburgh, Royal Observatory, Blackford Hill, Edinburgh, EH9 3HJ, UK}
\email{bbiller@ed.ac.uk}

\author[orcid=0000-0002-9962-132X,sname='Sutlieff']
{Ben J. Sutlieff}
\affiliation{Institute for Astronomy, University of Edinburgh, Royal Observatory, Blackford Hill, Edinburgh, EH9 3HJ, UK}
\email{ben.sutlieff@roe.ac.uk}

\author[0009-0008-2252-7969]
{Klaus Subbotina Stephenson}
\affiliation{University of Victoria, 3800 Finnerty Rd Victoria, BC V8P 5C2, Canada}
\affiliation{Herzberg Astronomy and Astrophysics Research Centre, 5071 W. Saanich Rd, Victoria, BC, V9E 2E7, Canada}
\email{klausss@uvic.ca}

\begin{abstract}

We report the direct imaging detection of a stellar companion to HD 101452 (HIP 56925; Gaia DR3 5384905720847544192), an A-star historically used as an infrared flux calibrator. The companion was identified from an asymmetry in the stellar PSF using short-integration ($<1$ minute), noncoronagraphic JWST MIRI imaging from 15 - 25.5$\mu$m that was obtained as part of the absolute flux calibration program CAL 4496. We detect the stellar companion in four MIRI imaging filters at a projected separation of 1.3\arcsec{} with a mid-infrared flux ratio of approximately $27\times$ warranting its removal from the ensemble of JWST flux calibrator systems. This detection also demonstrates that noncoronagraphic MIRI imaging can recover companions at separations below $3\lambda/D$ when a reference star with a closely matched flux level to the science target is available. Using the measured contrast curves from the calibration data, we evaluate the predicted sensitivity achievable with similar short MIRI observations for nearby stars ($<30$ pc) that are of interest to the direct imaging community with comparable brightness to HD 101452 ($W4\sim6.8$\,mag). We find that integrations of only 17--42 s are be capable of detecting sub-Jupiter-mass planets at solar system-like separations around a subset of neighboring systems if the PSF subtraction is conducted with closely flux-matched references. These results demonstrate that minutes-length, noncoronagraphic MIRI observations can provide a powerful and efficient new avenue for exploring the cold giant planet population around nearby stars.

\end{abstract}


\section{Introduction \label{sec:intro}}

 The JWST absolute flux calibration programs establish consistent, high-precision photometry across the observatory's instruments. Achieving the targeted flux accuracy of better than 5\% requires a carefully vetted ensemble of calibration stars that are free from nearby companions and disks \citep{Gordon2025}. However, JWST’s exceptional infrared sensitivity and angular resolution can reveal faint astrophysical sources that are inaccessible to vetting efforts at other facilities.  These vetting challenges are particularly acute in the mid-infrared, where the MIRI imager improves substantially upon the sensitivity of earlier ground- and space-based observatories at wavelengths of 5.6--25.5 $\mu$m. A variety of accidental discoveries have already been made in JWST images intended for another purpose \citep[e.g.][]{Burdanov2025, Sun2022, Venner2025}.

 A-star calibrators are especially susceptible to contamination from stellar companions and circumstellar dust that can influence the photometric calibration or distort the point-spread function (PSF) shape. Approximately one-quarter of A stars host a close stellar companion \citep{DeFurio2025} and one-third exhibit a 24$\mu$m excess indicative of a debris disk \citep{Su2006}. Vega, which historically defined the photometric magnitude scale, is itself surrounded by a prominent debris disk that is readily detected with JWST MIRI \citep{Su2024}. Recent MIRI observations of nearby systems such as $\epsilon$ Eri \citep{Wolff2025}, Fomalhaut \citep{Gaspar2023}, and TWA~7 \citep{Lagrange2025} further demonstrate the instrument's ability to reveal cold circumstellar material and faint companions in the mid-infrared.

HD 101452 (HIP 56925; Gaia DR3 5384905720847544192) is an A2V star \citep{Houk1982} that has served as an infrared flux calibrator for more than four decades \citep{Elias1982, Allington-Smith2007, Carter1990, Engelbracht2007, Rieke2008, Gordon2025}. Most recently, it was included as a JWST absolute flux calibrator in the CAL 4496 program \citep{cal4496, Gordon2025}. 
However, several observations suggest that \Lumpford{} hosts a companion which may make it unsuitable as a flux calibration source for JWST and future large aperture telescopes.

 In JWST MIRI imaging data, the source exhibits a distinctive asymmetric lump  embedded in the stellar PSF (see Figure \ref{fig:MIRIimgs}). 
 An analysis of the Rocky Worlds DDT program with the SPARTA pipeline measured an anomalously high infrared excess of 6\% at 15\,$\mu$m \citep{Xue2025}. \cite{Kouwenhoven2005} state that they recognize HD 101452 as a binary star system but do not expand on the properties of the noted companion. 
\Lumpford{} is also noted as a binary in the Fourth Catalog of Interferometric Measurements of Binary Stars \citep{Hartkopf2001}.\footnote{Fourth Catalog of Interferometric Measurements of Binary Stars: \url{https://astro.gsu.edu/wds/int4.html} with a reference code `HIP1997g,' indicating a non-component double star solution (MultFlag=G; i.e., acceleration solution) in the \textit{Hipparcos} catalog \citep{Perryman1997}}
Together, these indicators of multiplicity make HD 101452 a compelling target to perform a high-contrast imaging analysis aimed at determining whether contaminating flux warrants its removal as a JWST calibrator.

Because \Lumpford{} has an apparent brightness comparable to nearby stars targeted by other direct-imaging surveys (W4 = 6.8\,mag; \citealt{Wisemags}), the flux calibration program observations also provide a useful test case for evaluating the high-contrast imaging performance of short MIRI integrations. 
Detecting frigid giant planets at near-infrared wavelengths can be challenging because water clouds and enhanced metallicity may strongly suppress $3.5-5.5\mu$m flux \citep{Lacy2023, Matthews2024, Bowens-Rubin2025NircamYellsAtCloud, BardalezGagliuffi2025, Matthews2026, Sanghi2026, Sanghi2026epseri}.  Thus, there has been a growing effort in the exoplanet direct-imaging community to understand how to best use JWST's mid-infrared capabilities beyond 15 micron to detect planets using the wavelengths that are more agnostic to a planet's composition and atmospheric properties. 


Several JWST programs are currently demonstrating the power of using noncoronagraphic MIRI imaging for studying frigid giant planets and disk structures near and beyond their system's snow lines. The GO 6122 \citep{CoolKidsprop} and SURVEY 8581 \citep{HOTHprop} programs are using F2100W imaging to search for companions colder than 125\,K around the nearest star systems ($<8$\,pc) and recently showed that this observing mode is capable of detecting planets colder than Saturn in the very nearest systems \citep{Bowens-Rubin2025NircamYellsAtCloud}. Similar MIRI imaging completed by SURVEY 3964 \citep{Poulsen2023}, SURVEY 4403 \citep{MEOW}, and GO 7833 \citep{PAWS} have been critical for identifying giant planet candidates orbiting white dwarfs to understand the planet population around post-main-sequence stars (see \citealt{Mullally2024, Limbach2024, Albert2026}).  
\cite{Sanghi2026} used approximately one hour of F1800W, F2100W, and F2550W imaging from GO 8714 to study the composition of Eps Indi Ab and found evidence that the mid-infrared 25.5$\mu$m photometry is consistent with the \cite{Lacy2023} water-cloud models.  \cite{Li2026} repurposed long-duration F1500W time-series observations obtained as part of the GO 3730 Hot Rocks transit program \citep{HotRocksprop} to perform a deep high-contrast imaging search for outer companions.   
The on-sky performance of very short ($<5$ minute) noncoronagraphic MIRI observations for high-contrast imaging has yet to be explored.

In this work, we present a high-contrast imaging analysis of the HD 101452 MIRI observations from CAL 4496 to assess the system's suitability as a JWST flux calibrator and to evaluate the performance of short-integration noncoronagraphic MIRI imaging for directly detecting cold sub-Jupiter mass exoplanets. Section \ref{sec:obs} describes the observational setup used by the JWST absolute flux calibration programs to observe HD 101452 and the reference star. 
 Section \ref{sec:reduction} details the PSF-subtraction analysis performed with \texttt{VIP}. In Section \ref{sec:results}, we measure the position and flux of the detected companion, confirm its proper motion, and classify the companion. In Section \ref{sec:discussion}, we highlight the implications for future calibrator target selection and evaluate how short-exposure MIRI high-contrast imaging may perform on observations of nearby stars. Section \ref{sec:conclusion} overviews our primary takeaways.


\section{OBSERVATIONS \label{sec:obs}}

The MIRI observations used throughout this work of HD 101452 were obtained on Jun 22 2024 UTC as part of the JWST absolute flux calibration program CAL 4496 \citep{cal4496}. This program observed A-star calibrators across multiple JWST instrument modes and filters. 

The system was observed with the MIRI imager using the \texttt{brightsky} subarray with the FASTR1 readout pattern in four filters (F1500W, F1800W, F2100W, and F2550W). The total integration times were short, ranging from 17.3\,s -- 41.5\,s with 5--12 groups per integration (F1500W: 17.3s; F1800W: 34.6s; F2100W: 34.6s; F2550W: 41.5s). 
A four-position large-cycle dither pattern was used for each filter.

We adopted a second A-star from CAL 4496 (HD 2811) as the PSF reference for the flux subtraction. This star has a similar mid-infrared brightness to \Lumpford{}, although it is slightly fainter (W4=7.0\,mag; \citealt{Wisemags}). The observations used the same subarray, filters, and dither pattern as \Lumpford{} with slightly longer imaging integration times with 6--20 groups to yield a consistent overall flux between the calibration targets in the CAL 4496 program (F1500W: 20.8\,s; F1800W: 34.6\,s; F2100W: 34.6\,s; F2550W: 69.2\,s). 

To diversify the set of PSF references, we incorporated additional HD 2811 observations into the PSF reference library from CAL 6604 (15 Oct 2024 UTC; \citealt{cal6604}) and CAL 7487 (30 September 2025 UTC; \citealt{cal7487}). These images shared the same observing setup as the images of HD 2811 from CAL 4496.

\section{DATA REDUCTION \label{sec:reduction}}

The images used in this analysis were retrieved using the MAST archive.\footnote{MAST: \url{https://mast.stsci.edu/portal/Mashup/Clients/Mast/Portal.html}}
 The observations are publicly available through doi: 10.17909/675g-g709.

To subtract the stellar PSF, we followed the high-contrast imaging procedures developed for the GO 6122 MIRI observations presented in \cite{Bowens-Rubin2025NircamYellsAtCloud}. We first applied a custom flat-field correction with the 
\texttt{MAGIC}\footnote{\texttt{MAGIC}: \url{https://github.com/kevin218/Magic}} 
Python package to improve background sensitivity using Version 2.0 with JWST pipeline version 3.0.0. 
We performed the PSF subtraction with the principal component analysis (PCA) routines implemented in the \emph{VIP: Vortex Imaging Processing} package (\texttt{VIP}) Version 1.6.4 \citep{VIP}. 

The images were first cropped to $240\times240$ pixels and corrected with \texttt{cube\_correct\_nan}. We then measured the stellar centroid in each frame using a custom two-dimensional Gaussian fitting routine applied to the central $15\times15$ pixels, and the images were aligned to a common center with the pandas \texttt{shift} function. 
We performed the PSF subtraction using a reference differential imaging strategy with the \texttt{PCA} algorithm with \texttt{pc=3} and no center masking. 

\begin{figure*}
    \centering
    \includegraphics[width=1\linewidth]{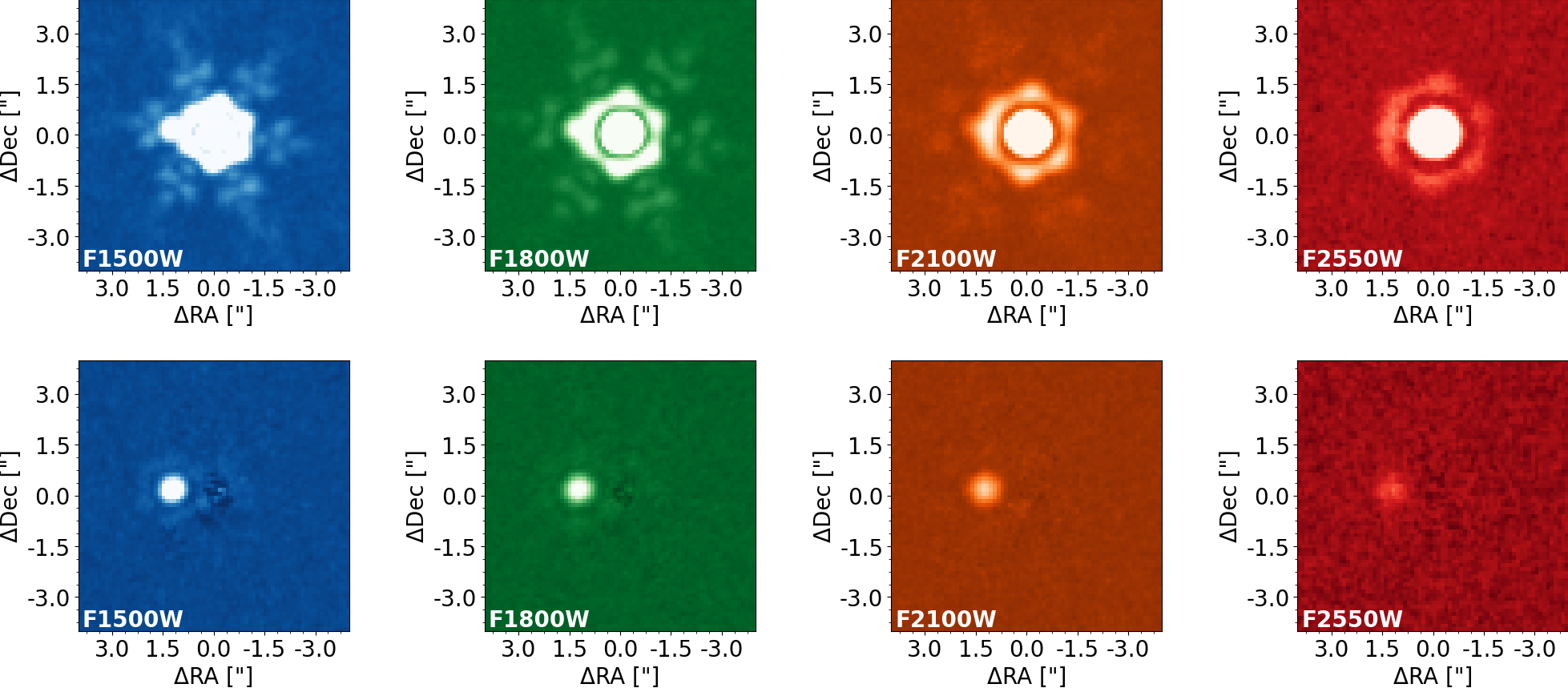}
    \caption{\textbf{JWST MIRI images of the HD~101452 system before and after PSF subtraction.} The top row shows 100$\times$100 pixel cutouts of the MIRI images centered on HD~101452 prior to PSF subtraction. The bottom row shows the corresponding PSF-subtracted images. All panels use a common surface brightness scale (-5 to 50 MJy/sr). The companion is detected in all filters at a consistent location with a measured separation of $r = 1.276 \pm 0.004''$ and position angle $\theta = 84.3 \pm 0.2^{\circ}$ East of North at the time of observation.}
    \label{fig:MIRIimgs}
\end{figure*}

Figure \ref{fig:MIRIimgs} presents the MIRI images before and after PSF subtraction.
We detect a source consistent with a stellar companion at the same location in all four filters.  
The highest-significance detection occurred in the F1800W observations where the companion is recovered with a signal-to-noise ratio (S/N) of 40 as measured with VIP's \texttt{detection} function using 34s of integration time.

We then measured the astrometry and photometry of the source using the \texttt{mcmc\_negfc\_sampling} function built into \texttt{VIP}  following the steps from the tutorial.   
The Markov Chain Monte Carlo (MCMC) run was initiated using the position and flux values produced by the \texttt{detection} and \texttt{first\_guess} functions with 150 walkers and a maximum of 1500 iterations.  
Table \ref{tab:results} summarizes the companion measurements derived from this MCMC analysis.

   We measure the source's separation to be $r=11.603\pm0.037$\,pixels at the time of observation using a weighted mean across the four filters. Adopting the parallax measurement from Gaia ($6.05 \pm 0.06$\,mas; \citealt{Gaia2016a, GaiaDR3})
in conjunction with the MIRI imaging plate scale of 0.11 arcsec/px\footnote{JDocs reference for MIRI platescale: \url{https://jwst-docs.stsci.edu/jwst-mid-infrared-instrument/miri-observing-modes/miri-imaging\#gsc.tab=0}}, this corresponds to a projected separation of $1.276 \pm 0.004$ arcsec  (210.9 $\pm$ 2.2\,AU). 
 
  The frames were North aligned by rotating the images by the \texttt{PA\_V3} telescope roll angle as listed in the science frame's fits header (112.318$^{\circ}$) plus the MIRI instrument rotational offset of 4.835$^{\circ}$. 
Applying this rotation, we derive a position angle of the source of $\theta = 84.3 \pm 0.2^{\circ}$ east of north at the time of observation.

\begin{deluxetable}{lllllllllrl}
\tablecaption{Summary of Results \label{tab:results}}
\tabletypesize{\scriptsize}
\tablehead{
\colhead{Filter} &
 \colhead{Tot. Int.} & 
 \colhead{Host Meas. Flux} &
 \colhead{Host App. Mag} & 
 \colhead{Comp. App. Mag} & 
 \colhead{Comp. Abs. Mag} & 
 \colhead{$r$ (pixel) }& 
 \colhead{$r$ (arcsec) }& 
 \colhead{$\theta$ (deg.) }& 
 \colhead{S/N} & 
 }
 \startdata
F1500W & 17.3s & 31162 $\pm$ 623 $\mu$Jy & 6.83 $\pm$ 0.02 & 10.41 $\pm$ 0.02 & 4.32  $\pm$ 0.03 & 11.60 $\pm$ 0.05 & 1.276 $\pm$ 0.005 & 84.2 $\pm$ 0.2 & 37.8 \\ 
F1800W & 34.6s & 21936 $\pm$ 439 $\mu$Jy & 6.82 $\pm$ 0.02 & 10.37 $\pm$ 0.02 & 4.31  $\pm$ 0.03 & 11.61 $\pm$ 0.06 & 1.277 $\pm$ 0.007 & 84.4 $\pm$ 0.3 & 40.4 \\ 
F2100W & 34.6s & 16979 $\pm$ 340 $\mu$Jy & 6.79 $\pm$ 0.02 & 10.34 $\pm$ 0.02 & 4.25  $\pm$ 0.03 & 11.62 $\pm$ 0.13 & 1.278 $\pm$ 0.014 & 84.3 $\pm$ 0.6 & 31.7 \\ 
F2550W & 41.5s & 10820 $\pm$ 216 $\mu$Jy & 6.85 $\pm$ 0.02 & 10.40 $\pm$ 0.02 & 4.31 $\pm$ 0.03 & 11.43 $\pm$ 0.63 & 1.258 $\pm$ 0.069 & 85.2 $\pm$ 3.4 & 24.3 \\ 
Combined & -- & -- & -- & -- & -- & 11.60 $\pm$ 0.04 & 1.276 $\pm$ 0.004 & 84.3 $\pm$ 0.2 & \multicolumn{1}{l}{--} \\ 
\enddata
\tablenotetext{}{\textit{(Column 2)} The total integration times were retrieved at \url{https://www.stsci.edu/jwst-program-info/download/jwst/pdf/4496/}. \textit{(Column 3)} The measured flux of the host star was computed after the subtraction of the flux of the companion.
\textit{(Column 4 \& 6)} The apparent magnitude of the host star and the absolute magnitude of the companion are listed in the Vega magnitude system adopting the Zero Point values from the SVO Filter Profile service.\footnote{The SVO Filter Profile service was accessed at this website in May 2026: \url{https://svo2.cab.inta-csic.es/svo/theory/fps3/index.php?id=JWST}} The absolute magnitude was computed from the apparent magnitude assuming a distance of 165.232\,pc \citep{GaiaDR3}.  \textit{(Column 10)} The S/N was found using the max S/N output from the \texttt{detections} function within \texttt{VIP}. }

\end{deluxetable}


\section{Results \label{sec:results}}

\subsection{Contrast Performance \label{sec:cc}}

To quantify the sensitivity of the observation, we generated contrast curves using the \texttt{contrast\_curve} function in \texttt{VIP} with the same configuration adopted in the main analysis (\texttt{algo=pca}, no central masking). This function uses an injection-recovery based method to calculate the image noise and algorithm throughput to produce a 
contrast measurement as a function of radius from the host star. 
We adopt a Student's t-correction in order to   account for small sample statistics at the tightest separation angles as per the recommendations stated in \citealt{Mawet2014} under the conservative assumption that the data are speckle limited.

\begin{figure*}
    \centering
    \includegraphics[width=1\linewidth]{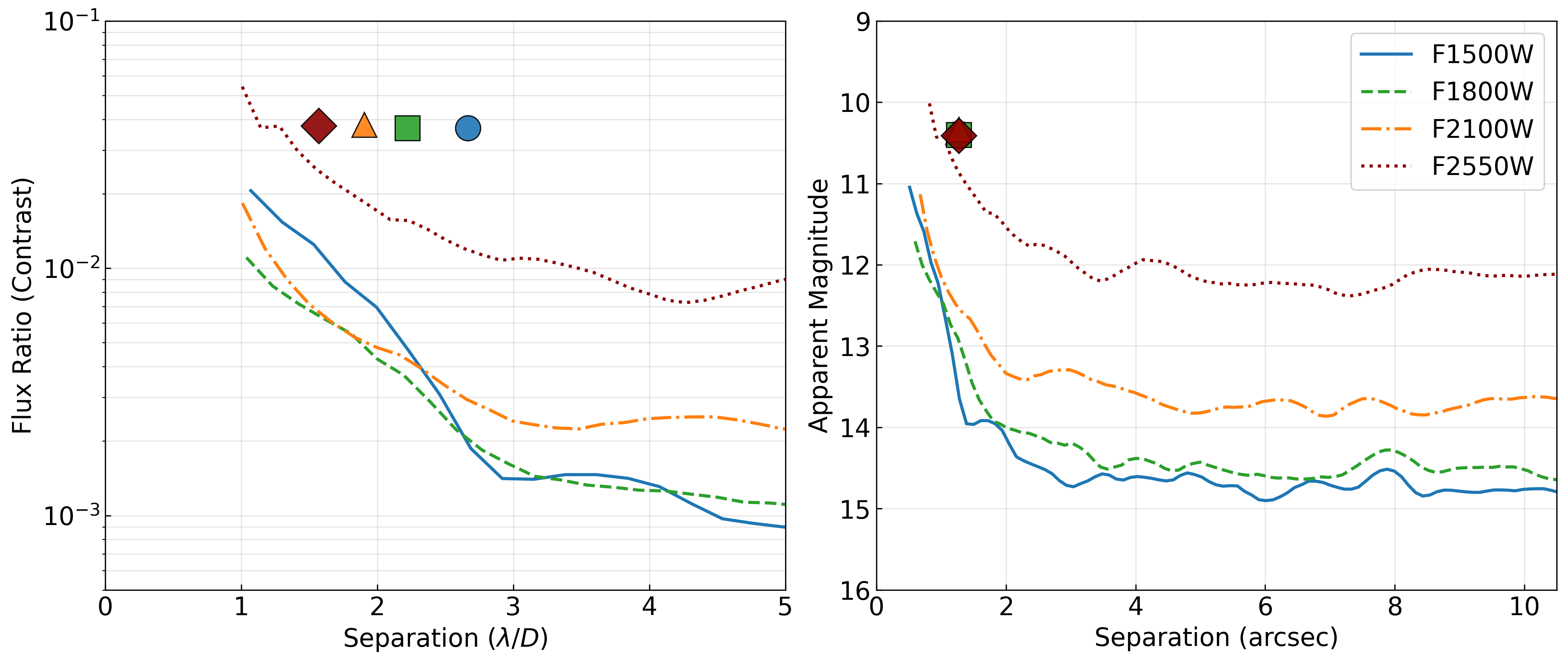}
    \caption{\textbf{Contrast Curves and Detected Companion}.\textit{(Left)} The 5 S/N contrast is plotted as the flux ratio as a function of the angular separation in units of $\lambda/D$, where $\lambda$ is the central wavelength of the filter and $D =6.5$m is JWST's primary mirror diameter.
     The photometry values measured for the companion are represented in each filter as the F1500W blue circle, F1800W green square, F2100W golden triangle, and F2550W red diamond. The errorbars for the photometry are plotted but lie inside the bounds of the symbol in each cases. 
    \textit{(Right)} The same 5 S/N contrast curves are shown in apparent Vega magnitude and plotted as a function of separation in arcseconds. 
    The stellar companion was detected in all four MIRI filters, including at the longest wavelengths where the projected separation is less than $2\lambda/D$. These observations demonstrate that noncoronagraphic MIRI imaging can recover companions at separations traditionally associated with coronagraphy provided that a well matched reference star  in flux counts is available for PSF subtraction.}
    \label{fig:cc}
\end{figure*}

To ensure an accurate estimate of the noise floor, we first removed the contribution of the detected companion using the \texttt{cube\_planet\_free} routine. This step used the companion position and flux measured from the MCMC analysis presented in Table \ref{tab:results}.
Figure \ref{fig:cc} shows the resulting contrast curves for the four filters. 

 The best 5 S/N flux ratio contrast was achieved using the 17s of F1500W 
 (5.9e-04; $\Delta$mag = 8.1).
The performance within $<2.5$ arcsec was not significantly degraded by the brighter/fatter residuals that previously limited the GO 6122 reductions in 
 \citealt{Bowens-Rubin2025NircamYellsAtCloud}. 
 This demonstrates that it is possible to mitigate the brighter/fatter effects in MIRI noncoronagraphic imaging inside $3 \lambda/D$ when a well-balanced reference star in flux level is available.

\subsection{Updated Photometry of the Host Star \label{sec:hostphot}}

We measured the disentangled photometry of the \Lumpford{} host star using the companion-subtracted images described in Section \ref{sec:cc} using aperture photometry.
The aperture radii and background annuli were taken from the CRDS encircled-energy reference file \texttt{jwst\_miri\_apcorr\_0014.fits}. The fluxes were computed using the seven encircled-energy aperture definitions provided in this file. The final flux in each band was taken to be the mean of the background-subtracted aperture measurements across these apertures.
We converted flux densities from janskys to magnitudes using the JWST MIRI photometric calibration. 
Specifically, we adopted the AB-Vega magnitude offsets from the CRDS reference file \texttt{jwst\_miri\_abvegaoffset\_0002.asdf} to convert from AB magnitudes to Vega magnitudes.

Because the source has very high signal-to-noise, the uncertainty in the photometry is dominated not by photon noise but by the absolute calibration of MIRI. We therefore adopt a 2\% systematic uncertainty on all fluxes, consistent with the current calibration precision \citep{Gordon2025} and propagate this to the reported magnitudes. These revised host star fluxes are reported in Table \ref{tab:results}.

\subsection{Common Proper-motion Confirmation of the Companion}

We first assess the likelihood that the detected source is a chance alignment with a background object using the \texttt{SIDERIS}\footnote{SIDERIS: \url{https://github.com/aiza-kenzhebekova/SIDERIS}} python package. This package allows the user to input the measured MIRI apparent magnitude and angular separation of a candidate to calculate the false probability rate as compared to the source density of the SMILES extragalactic survey \citep{Alberts2024, Zhu2026, Rieke2024}. We find a false-positive probability of 0.005\% for the source based on the F1500W and F2100W apparent magnitude and separation reported in Table \ref{tab:results}, indicating this source strongly favors a companion interpretation. 

The Gaia DR3 catalog from 2016 provides a resolved optical detection of a source at the same relative sky position as the source near \Lumpford{} (Gaia DR3 5384905720851582848). We use the Gaia DR3 catalog RA and Dec measurements to perform a common proper motion test, shown in Figure \ref{fig:cpm}. To assess the relative strength of the background and comoving hypotheses, we compute  $\chi^2$ for a model of common proper-motion ($\chi^2_\mathrm{CPM}$) and a background source ($\chi^2_{\mathrm{BG}})$ model. The comoving scenario is overwhelmingly favored with $\Delta \chi^2 = \chi^2_\mathrm{BG} - \chi^2_\mathrm{CPM} = 5953$, strongly supporting a bound interpretation. We therefore conclude that the source discussed throughout this work is a bound companion to HD 101452. 



\subsection{Classification of the Companion}

\begin{figure*}
    \centering
    \begin{subfigure}[t]{0.49\linewidth}
        \centering
        \includegraphics[width=\linewidth]{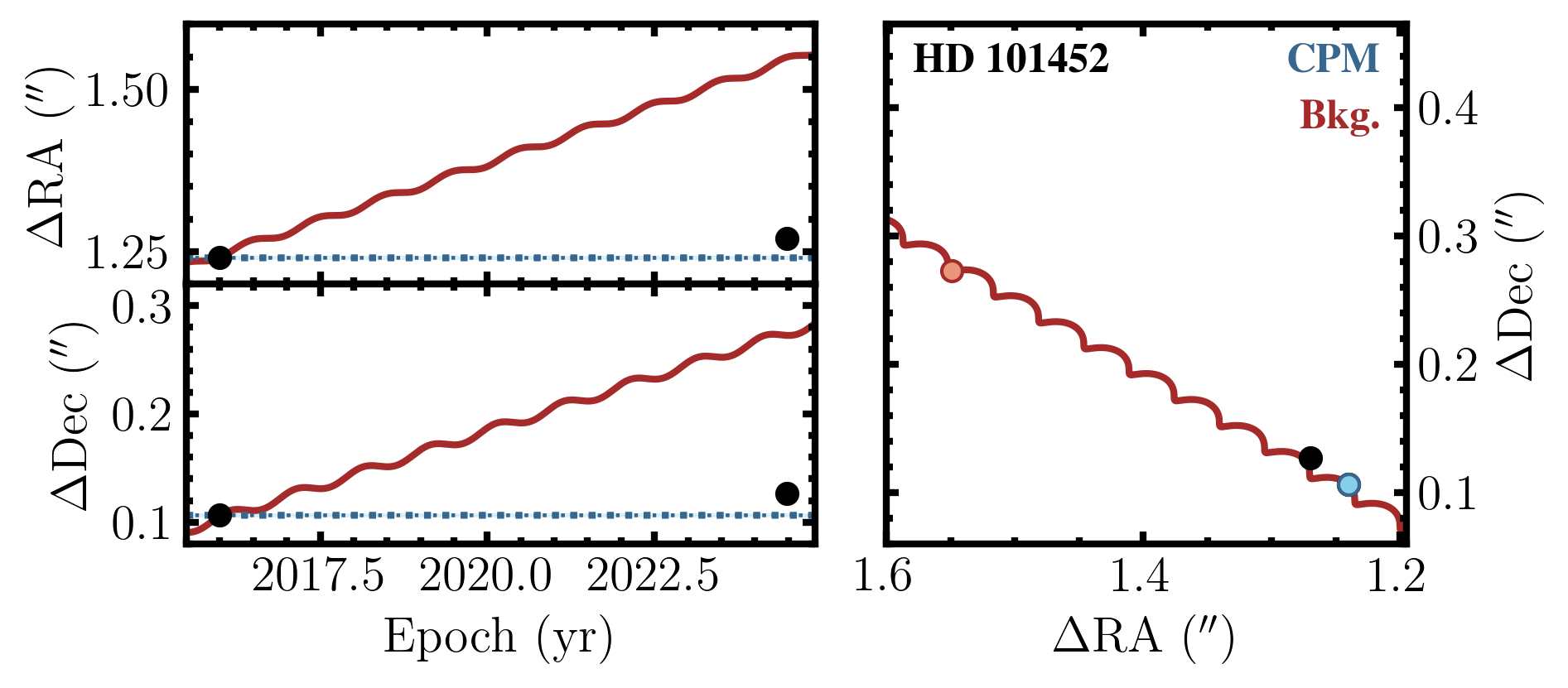}
        \caption{Common proper motion verification}
        \label{fig:cpm}
    \end{subfigure}
    \hfill
    \begin{subfigure}[t]{0.5\linewidth}
        \centering
        \includegraphics[width=\linewidth]{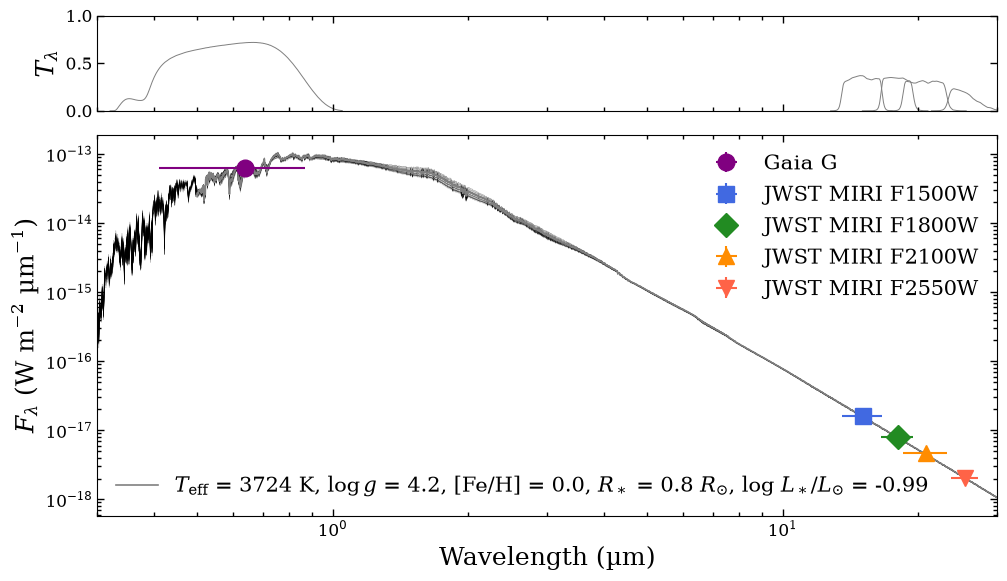}
        \caption{Spectral energy distribution fit}
        \label{fig:SED}
    \end{subfigure}

    \caption{\textbf{Common proper-motion verification and source identification.}
    \textbf{(a)} The relative astrometry of the detected source between the Gaia DR3 information and the MIRI detection is consistent with a comoving companion (dotted blue line) over the expected motion of a stationary background object (solid red line). 
    This test confirms the source is gravitationally associated with the primary star with the comoving model preferred by $\Delta \chi^2 = 5953$. \textbf{(b)} An SED fit of the companion was made using the \texttt{species} package incorporating the Gaia DR3 G magnitude \citep{GaiaDR3} and the four MIRI photometry points listed in Table \ref{tab:results}.  The best-fit SED is consistent with a star with a classification between K7V to M0.5V \citep{Pecaut2013}.}
    \label{fig:companion_verification}
\end{figure*}

To classify HD 101452 B, we fit the companion's spectral energy distribution (SED) using the \texttt{species} Python package (version 0.10.5). We adopt the assumption that the source is a single main-sequence star without a circumstellar disk. We incorporate the four MIRI photometric measurements listed in Table~\ref{tab:results} together with the companion's Gaia DR3 G magnitude ($14.02 \pm 0.04$) \citep{GaiaDR3} to perform a fit to the Phoenix NewEra atmospheric models \citep{Hauschildt2025}. The parallax was constrained to the Gaia DR3 measurement ($6.0521\pm0.055$ mas), while broad priors were adopted for the temperatures (2500 -- 10,000 K), radius (6--10 $R_{\rm Jup}$), $\log g$ (3.5--5.0), and [Fe/H] ($-1$ -- $+1$).
The best-fit SED is shown in Figure~\ref{fig:SED}. The fit yields an effective temperature of $3724^{+66}_{-67}$ K, 
 luminosity of $\log(L/L{\odot})=-0.99\pm0.02$, 
  radius of $0.76 \pm 0.01 R_{\odot}$ ($7.5 \pm 0.1 R_{\rm Jup}$), 
and metallicity of [Fe/H]$=-0.03\pm0.33$. 

These parameters do not correspond to a single spectral type in the empirical relations of \citealt{Pecaut2013} but lie near the boundary between the M and K stellar classifications. If the Gaia G mag is used as an anchor for the spectral fitting, the mid-infrared photometry appears in slight excess to the tracks as stated in \citealt{Pecaut2013}.\footnote{Classification tables for \citealt{Pecaut2013} were accessed online at \url{https://www.pas.rochester.edu/~emamajek/EEM_dwarf_UBVIJHK_colors_Teff.txt}} It is uncertain if this excess could be caused by an astrophysical source such as an unseen disk or third companion or if the Gaia G-band photometry could have a systematic error causing it to be slightly underreported due contamination from the light from the nearby host star.
We therefore conservatively conclude that the imaged companion is consistent with a late K dwarf to early M dwarf (K7V--M0.5V).

\section{Discussion \label{sec:discussion}}

\subsection{Implications for Future Calibrator Target Selection \label{sec:futurecal}}

The extreme flux contrasts achieved by current and future high-contrast imaging instruments require increasingly rigorous characterization of calibration targets for both photometric calibration and PSF subtraction. Consequently, robust pre-mission target vetting has become essential. Similar challenges will soon be faced by the Roman Coronagraph Instrument \citep{Hom2026}, as well as the ELTs and the HWO.

The existence of a stellar companion to \Lumpford{} was not definitively recorded in the literature prior to this work. However, \citet{Kouwenhoven2005} allude to the detection of a stellar companion, and we propose that this likely represents a predetection of the same stellar companion reported here. 
This demonstrates that comparatively benign stellar companion detections may have unexpected significance for planning of future observations and emphasizes the importance of duly reporting such detections in the literature.

The rapid expansion of all-sky surveys has dramatically improved our sensitivity to identify stellar companions to bright stars.
In particular, \textit{Gaia} \citep{Gaia2016a} provides a powerful tool for vetting calibration targets.
Independent of the resolved detection of the companion in \textit{Gaia}, the available astrometric data of the host star provides strong evidence that \Lumpford{} is a multiple star system. 
The \textit{Gaia}~DR3 astrometric solution for \Lumpford{} has a renormalized unit weight error (RUWE) of 2.309 \citep{GaiaDR3}, which is far above the RUWE > 1.25 heuristic for astrometric binary detection in \textit{Gaia}~DR3 suggested by \citet{Penoyre2022}.
Furthermore, an astrometric acceleration solution was reported in the \textit{Hipparcos} catalog \citep{Perryman1997}, and inspection of cross-calibrated \textit{Hipparcos-Gaia} astrometry reveals a highly significant proper-motion anomaly of $\Delta\mu\approx4.2$~mas~yr$^{-1}$ \citep[$\chi^2=2582$, or $51\sigma$;][]{Brandt2021HGCA}, corresponding to a change in tangential velocity of approximately 3300~m~s$^{-1}$ at the distance of the system.
Given the wide projected separation of the resolved companion ($210.9\pm2.2$ AU), this may even suggest that these astrometric signatures indicate the presence of an additional, shorter-period third object in the system.
Had these diagnostics been available during the selection of JWST calibration targets, they may have provided sufficient grounds to exclude \Lumpford{}. 

\subsection{Exoplanet Sensitivity if Scaled to Neighboring Systems} 

Although \Lumpford{} is an A2V star at 165\,pc \citep{GaiaDR3}, its apparent magnitude (W4 = 6.8\,mag; \citealt{Wisemags}) is comparable to a set of nearby ($<30$\,pc) stars of lower mass of interest to the direct-imaging community. We therefore use the CAL 4496 observations as an empirical reference point to estimate the sensitivity that similar short-integration MIRI observations could achieve for identifying companions orbiting closer, lower-mass host stars.

To translate the measured contrast performance into planetary mass sensitivity, we used the \texttt{bex-atmo2023-ceq} evolutionary models \citep{Linder2019, atmo2023, Marleau2019} as implemented in the \texttt{madys} package.  This model grid is among the few available that extends to the planetary mass regime probed by these observations and is therefore well suited to this analysis. However, the BEX-ATMO models assume a clear atmosphere with solar metallicity ([Fe/H] = 0) and C/O = 0.55. These assumptions introduce an important model dependence into the inferred mass sensitivity. In particular, the abundance of CO$_2$ can affect planetary emission near 15$\mu$m and variations in metallicity and cloud properties can shift the inferred detectability by up to $\sim 1 M_{\rm Jup}$ \citep{Mang2026}. While the mid-infrared variation is significantly less than the variation in the near-infrared from 3-5$\mu$m, the mass limits presented here should nonetheless be interpreted as model-dependent estimates rather than comprehensively atmosphere-independent detection limits.

Teegarden's Star is a mature M7.5 dwarf \citep{Fuhrmann2012} in the 25th closest neighboring system (3.8\,pc; \citealt{GaiaDR3}) 
with multiple confirmed radial velocity planets \citep{Zechmeister2019}. This star was part of the target list for JWST SURVEY 8581 \citep{HOTHprop} although was not observed. 
Assuming an age of 8 Gyr \citep{Zechmeister2019}, the equivalent companion detection to the HD 101452 system would be in the mass regime of a low-mass brown dwarf (20-25 $M_{\rm Jup}$) at 4.8 AU.  Exoplanets with masses $<1M_{Jup}$ would be detectable in the F2100W filter beyond 4'' (15 AU) to masses as low as $0.75M_{\rm Jup}$.

We next consider a sample of nearby, young, M-dwarf systems (6.5–8\,pc; $\sim$700Myr) that were targeted by GO 8826 using NIRCam coronagraphy \citep{Lawson2025prop} to study their disks. Five targets in this program had W4 magnitudes similar to \Lumpford{} (6.77–7.05\,mag): G 13-22, GL Vir, G 227-22, SCR J0740-4257, and G 141-36. 
The detected companion would correspond to a low-mass brown dwarf (20-25 $M_{\rm Jup}$) at 8-10 AU.  The F2100W filter provides the best constraints reaching 5 S/N detections of planets to masses as low as $0.7M_{Jup}$.

We also examine the limits around two young moving group systems at distances $>15 pc$ that are members of the AB Dor moving group (age $\sim 100$ Myr  \citealt{Luhman2005}) that were observed using NIRCam coronagraphy to search for low-mass cold giant planets.  HIP 17695 (16.8\,pc; W4 = 6.5\,mag) is an M-dwarf that was observed as part of GTO 1184 \citep{Schlieder2017}. 
2MASS J07234358+2024588 (27.7\,pc;  W4 = 6.6\,mag) is a K-dwarf observed as part of SURVEY 6005 \citep{Biller2025}.  The companion detection in these systems is analogous to a high-mass brown dwarf (50--60$M_{\rm Jup}$). By this distances, the F2100W filter is no longer the optimal choice of filter, with F1500W and F1800W yielding better access to the lowest mass planets. Sub-Jupiter mass planet limits are achievable in the HIP 17695 system but not in the more distant 2MASS J07234358+2024588 system using this sort of short-integration setup.  This highlights that the largest advantages for this sort of short-integration MIRI imaging may lie within $<20$\,pc sample when being applied to study the sub-Jupiter mass giant planet population.

Short-integration MIRI imaging is unfortunately not sufficient to meet the needs of all science cases in the JWST direct-imaging programs that aim to study cold giant planets. 
For example, GJ 832 (5\,pc; 4.6\,Gyr) hosts a Jupiter-mass planet candidate that will be observed by the medium-sized GO 10758 program using the MIRI coronagraph. Even assuming the \Lumpford{} contrast performance can be achieved for this brighter star ($W4 = 4.0$\,mag), the inferred sensitivity would not reach \textbf{below $1.7\,M_{\rm Jup}$} at any separation. Nevertheless, it is remarkable that a noncoronagraphic MIRI observation requiring less than one minute of integration approaches the sensitivity expected from a dedicated medium-sized coronagraphic program. A broader exploration of the trade space between noncoronagraphic and coronagraphic MIRI observations, particularly at integration times comparable to those of GO 10758, could identify opportunities to improve the efficiency of future JWST observing programs.

Overall, we find that $<1$\,min MIRI imaging can provide the sensitivity to study the cold gas-giant planet population at solar-system-like scales (5-50\,AU). In favorable cases, these observations extend into the sub-Jupiter-mass regime. This sensitivity is achievable for nearly all of the nearest systems ($\lesssim4$\,pc) and for young systems within $\sim20$\,pc, although the precise mass limits depend on the combination of distance, system age, host-star brightness, adopted atmospheric models, and MIRI observing mode. 

Among the filters considered, F2550W provides the weakest sensitivity to the lowest-mass planets using short integrations because of its higher background noise relative to the shorter-wavelength MIRI filters.  A trade space of the F1500W, F1800W, and F2100W MIRI filters must be evaluated on a case-by-case basis  
when trying to optimize for the filter most likely to yield the highest S/N detections. The optimal filter choice may also shift in the case of longer integration times.

For such short integrations, telescope overheads dominate the total observing time. For example, the Astronomer Proposal Tool charges approximately 1\,hour to slew to a new target. Nevertheless, the results suggest that a medium-sized MIRI program could include $\sim$100 observations of nearby systems to place atmosphere-independent constraints on the population of cold gas-giant planets. Pairing short MIRI observations with programs already using another instrument would reduce the overhead associated with a separate slew, requiring only an instrument change ($\sim$30 minutes). Such pairings may prove especially valuable for NIRCam and NIRSpec studies of young disks where additional mid-infrared imaging could reveal planets responsible for shaping the disk morphology. 
For observations already using MIRI imaging, switching between MIRI imaging filters incurs little overhead ($<10$ minutes when no subframe change is required). Consequently, these results imply that it is best practice to obtain multiple MIRI colors whenever even modest scientific gains are possible to overall increase the efficiency of MIRI's JWST exoplanet and disk programs.

\begin{figure*}
    \centering
    \includegraphics[width=1\linewidth]{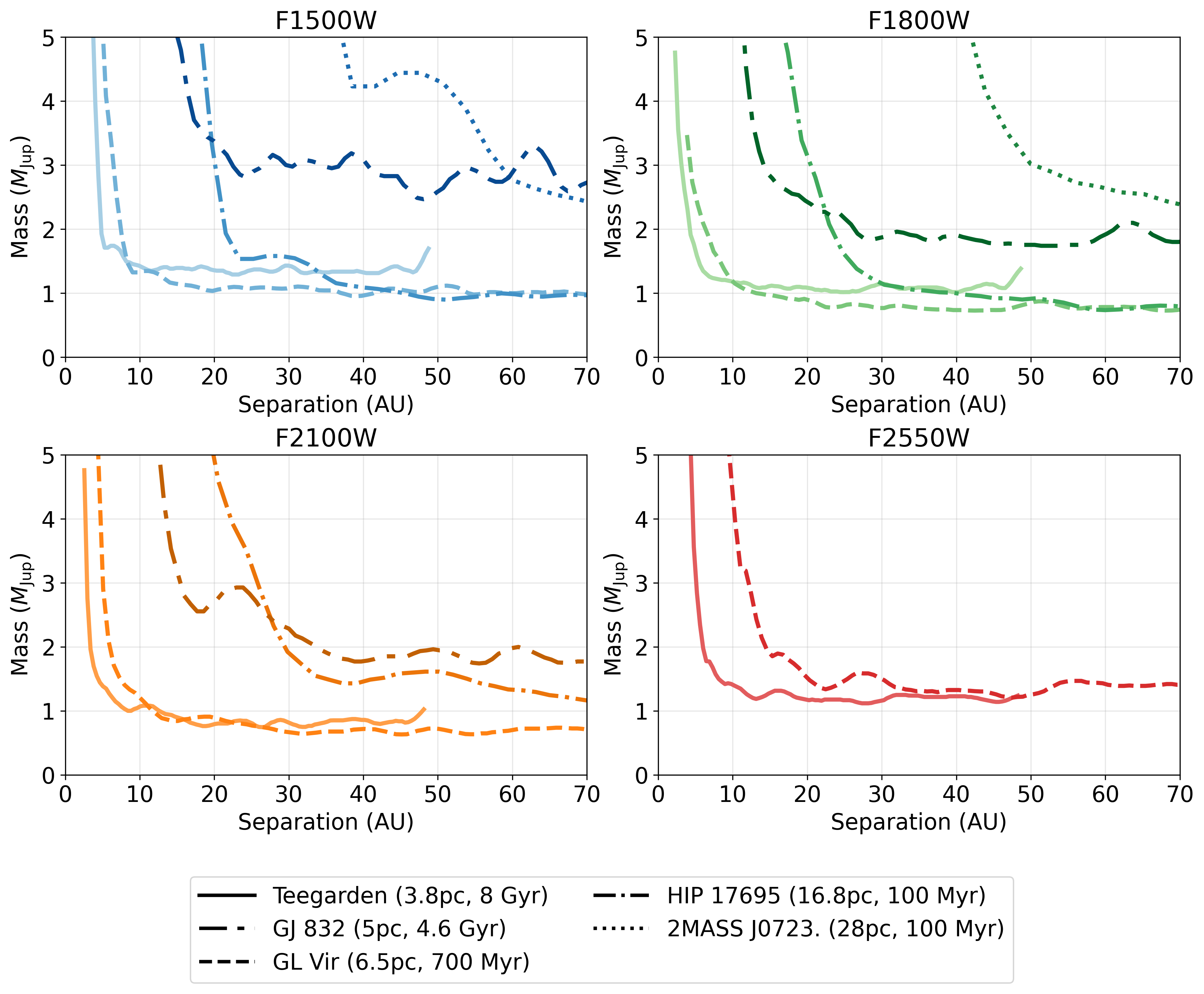}
    \caption{\textbf{S/N=5 mass limits for neighboring system examples.} The \Lumpford{} contrast curves reported in Figure \ref{fig:cc} were translated to mass limits using \texttt{madys} for five nearby systems. 
    These examples appeared on the target lists of one or more JWST direct imaging programs prior to Cycle 6, and most have a WISE W4 mag similar to \Lumpford{} ($W4 \sim 6.8$). These plots are displayed to 70 AU, but the contrast curves extend to wider separations that are not shown for some targets. F2550W is the least advantageous filter for cold planet hunting using this sort of short-integration MIRI imaging because of the high instrument background levels as compared to the other MIRI filters. The mass constraints of GJ 832, HIP 17695, and 2MASS J07234358+202458 are not shown in the F2550W plot because the limits are greater than 5$M_{jup}$ at all separations.  
     Sensitivity to detect sub-Jupiter-mass gas giant planet candidates can be reached using 34s of F1800W / F2100W for three of the five example cases where a combination of very near proximity ($<4$\,pc) or youth ($<100$\,Myr) is present.} 
    \label{fig:madys}
\end{figure*}

\section{CONCLUSIONS \label{sec:conclusion}}

Our analysis confirms that the infrared flux calibrator HD 101452 is a binary system. At the epoch of the JWST observations (2024 June 22 UT), the companion lay at a projected separation of $1.276 \pm 0.004$ arcsec (210.9 $\pm$ 2.2 AU) with a mid-infrared flux ratio of approximately $27\times$ relative to the primary. The companion introduces flux contamination that exceeds the current accuracy of the JWST absolute flux calibration \citep{Gordon2025} while simultaneously producing an asymmetric stellar PSF. We therefore recommend removing HD 101452 from the JWST calibration sample. Because this star is one among an ensemble of JWST calibration sources, we expect its removal to have minimal effects on the overall calibration of the observatory, but it may affect custom reductions that operate with a smaller number of flux calibrators.



We confirmed that the detected source is a bound stellar companion to \Lumpford{} through a common proper-motion analysis. The source was resolved by Gaia (Gaia DR3 538490572085158284), which provided an 8 yr time baseline with the JWST MIRI images to confirm its co-proper motion.   An SED fit using the measured photometry was consistent with the companion having a stellar classification between K7V and M0.5V. 

Beyond its impact on calibration, this system provides a demonstration of the capability of short-integration, noncoronagraphic MIRI imaging. The companion was detected at separations $<3\lambda/D$ showing that it is possible to mitigate the residuals caused by the brighter/fatter detector effect at the tightest separations when a closely matched reference star in flux is available.
This result demonstrates that MIRI integrations of only tens of seconds can access a region of parameter space previously associated primarily with coronagraphic observations.


For the nearest systems, similar short MIRI observations are sensitive to cold giant planets on solar system scales (5--50 AU). In favorable cases, this mode can detect planets below $1\,M_{\rm Jup}$. Because these observations require only tens of seconds of integration, a medium-sized JWST program could provide approximately 100 observations of nearby stars despite the $\sim$1\,hour slew overhead required for each independent visit. Even greater efficiency could be achieved by pairing short MIRI observations with already scheduled visits, reducing the additional overhead to approximately 30\,minutes for an instrument change. 
Rather than serving solely as a calibration mode, short-integration, noncoronagraphic MIRI imaging provides a powerful and efficient means of exploring the cold outer architectures of nearby planetary systems. Incorporating these observations into future JWST programs could substantially increase the mission's scientific return at only modest additional cost.


\begin{acknowledgments}

We would like to acknowledge the thousands of people who dedicated themselves to the design, construction, commissioning, and operation  of JWST. We elect to use the acronym of the telescope without using the spelling of the telescope's full name as per the AAS style guidelines in order to acknowledge the uncertainty of James Webb's involvement in the discriminatory firing of NASA employees during the Lavender Scare. 

R.B.R would like to thank Vito Squicciarini and Rapha\"{e}l Bendahan-West for their support with \texttt{madys} along with Aniket Sanghi for sharing his expertise in characterizing cold giant exoplanets.  
We thank the members of the University of Michigan Formation and Evolution of Planetary Systems
``FEPS'' group, which is led by Micheal Meyer, for their feedback during group meetings. R.B.R is also grateful for the support from members of the Endor discord server led by Mary Anne Limbach and the Vanderburgers discord server led by Andrew Vanderburg. 


This work used observations made with the NASA/ESA/CSA JWST as part of CAL 4496, ``Absolute Flux Calibration (A Dwarfs),'' in which Karl Gordon was the PI. We thank the PI and CoI's of CAL 4496 including Karl Gordon, Sheri Holfeltz, Greg Sloan, Charles Proffitt, and Kevin Volk. 

The data were obtained from the Mikulski Archive for Space Telescopes at the Space Telescope Science Institute, which is operated by the Association of Universities for Research in Astronomy, Inc., under NASA contract NAS 5-03127 for JWST. We also acknowledge the grant support of STScI grant
JWST-GO-06122.014. 


\end{acknowledgments}

\begin{contribution}
Rachel Bowens-Rubin identified the distinct PSF shape of \Lumpford\ from the MIRI data, performed the PSF subtraction, and drafted the majority of the manuscript.  Mary Anne Limbach performed the updated photometry of the \Lumpford\ host star, authored Section \ref{sec:hostphot}, and provided detailed early feedback on all sections. Alexander Venner was the primary author of Section \ref{sec:futurecal} and provided information about the stars astrometric acceleration.   Kyle Franson performed the common proper motion confirmation and created Figure \ref{fig:cpm}. Emily Pass and Andrew Vanderburg provided assistance to the stellar SED fitting, detailed corrections, and suggestions for improving the readability of early figure drafts.  Kevin Stevenson performed the reduction with \texttt{Magic} to produce the custom flat field correction. Aiza Kenzhebekova, Beth Biller, Klaus Subbotina Stephenson and Ben Sutlieff are the creators of the \texttt{SIDERIS} code that was used to calculate the false detection probability.  
All co-authors provided suggestions and discussions in service of improving the manuscript.

\end{contribution}

\facilities{JWST}

\software{\texttt{VIP: Vortex Imaging Processing python package} \citep{VIP}, \texttt{species} \citep{Stolker2020}, \texttt{madys} \citep{Squicciarini2022}, 
\texttt{astropy} \citep{astropy:2013, astropy:2018, astropy:2022}.
ChatGPT was used to improve the wording at the sentence level and to assist in coding.
          }

\bibliography{sample701}{}
\bibliographystyle{aasjournalv7}

\end{document}